\documentclass[lettersize,journal]{IEEEtran}
\usepackage{amsmath,amsfonts}
\usepackage{array}
\usepackage[caption=false,font=normalsize,labelfont=sf,textfont=sf]{subfig}
\usepackage{textcomp}
\usepackage{stfloats}
\usepackage{url}
\usepackage{verbatim}
\usepackage{graphicx}
\usepackage{cite}
\usepackage{caption}
\usepackage{cite}
\usepackage{graphics}
\usepackage{graphicx}
\usepackage{textcomp}
\usepackage{comment}
\usepackage{url}
\usepackage{makecell}
\usepackage{xcolor}
\usepackage{siunitx}
\usepackage{booktabs}
\usepackage{lscape}
\usepackage{lipsum}
\usepackage{mwe}
\usepackage{hyperref}
\usepackage[autostyle]{csquotes}

\usepackage[noend]{algpseudocode}
\usepackage{algorithm, float}
\usepackage{tikz}

\usepackage{multirow}
\usepackage{adjustbox}
\usepackage{threeparttable}

\def\BibTeX{{\rm B\kern-.05em{\sc i\kern-.025em b}\kern-.08em
    T\kern-.1667em\lower.7ex\hbox{E}\kern-.125emX}}
\usepackage{balance}

\begin{document}
\title{PACMAN: a framework for pulse oximeter digit detection and reading in a low-resource setting}

\author{Chiraphat Boonnag, Wanumaidah Saengmolee, Narongrid Seesawad, Amrest Chinkamol, Saendee Rattanasomrerk, Kanyakorn Veerakanjana, Kamonwan Thanontip, Warissara Limpornchitwilai, Piyalitt Ittichaiwong, and Theerawit Wilaiprasitporn, ~\IEEEmembership{Senior Member,~IEEE}
\thanks{This work was supported by IEEE HAC/SIGHT, the National Research Council of Thailand (N35A650037), PTT Public Company Limited, and The SCB Public Company Limited. 
\textit{(Corresponding author: Piyalitt Ittichaiwong; Theerawit Wilaiprasitporn)}}
\thanks{C. Boonnag is affiliated with Biomedical Informatics Center, Faculty of Medicine, Chiang Mai University, Thailand.}
\thanks{W. Saengmolee, N. Seesawad, A. Chinkamol, K. Thanontip, W. Limpornchitwilai, and T. Wilaiprasitporn are with Bio-inspired Robotics and Neural Engineering (BRAIN) Lab, School of Information Science and Technology (IST), Vidyasirimedhi Institute of Science \& Technology (VISTEC), Rayong, Thailand (e-mail: theerawit.w@vistec.ac.th)}
\thanks{S. Rattanasomrerk is with Medensy Co., Ltd., Bangkok, Thailand}
\thanks{K. Veerakanjana and P. Ittichaiwong are with Siriraj Informatics and Data Innovation Center, Faculty of Medicine Siriraj Hospital, Mahidol University, Bangkok, Thailand}
}

\markboth{Journal of \LaTeX\ Class Files,~Vol.~14, No.~8, August~2021}%
{Shell \MakeLowercase{\textit{et al.}}: A Sample Article Using IEEEtran.cls for IEEE Journals}

\maketitle

\begin{abstract}
In light of the COVID-19 pandemic, patients were required to manually input their daily oxygen saturation (SpO\textsubscript{2}) and pulse rate (PR) values into a health monitoring system—unfortunately, such a process trend to be an error in typing. Several studies attempted to detect the physiological value from the captured image using optical character recognition (OCR). However, the technology has limited availability with high cost. Thus, this study aimed to propose a novel framework called PACMAN (Pandemic Accelerated Human-Machine Collaboration) with a low-resource deep learning-based computer vision. We compared state-of-the-art object detection algorithms (scaled YOLOv4, YOLOv5, and YOLOR), including the commercial OCR tools for digit recognition on the captured images from pulse oximeter display. All images were derived from crowdsourced data collection with varying quality and alignment. YOLOv5 was the best-performing model against the given model comparison across all datasets, notably the correctly orientated image dataset. We further improved the model performance with the digits auto-orientation algorithm and applied a clustering algorithm to extract SpO\textsubscript{2} and PR values. The accuracy performance of YOLOv5 with the implementations was approximately 81.0-89.5\%, which was enhanced compared to without any additional implementation. Accordingly, this study highlighted the completion of PACMAN framework to detect and read digits in real-world datasets. The proposed framework has been currently integrated into the patient monitoring system utilized by hospitals nationwide.
\end{abstract}


\section{Introduction}

\label{sec:introduction}
\IEEEPARstart{T}{he} demand for hospitalization during the COVID-19 pandemic has put the healthcare system on the verge of breaking down.
As a result of the significant increase in inpatient hospitalization, many hospitals are experiencing bed shortages. 
During the COVID-19 outbreak, telemedicine may be an alternative as healthcare providers are using telemedicine for remote patient monitoring, especially for mild symptom patients \cite{mann2020covid}. 
Healthcare providers chose to use telemedicine solutions in order to minimize direct physical contact, reduce COVID-19 transmission, and provide continuity of care. 

A COVID-19 remote patient monitoring program in many countries \cite{alboksmaty2022effectiveness, gordon2020remote, hutchings2021virtual} recommended the use of home pulse oximetry monitoring to avoid overcrowding in hospitals. The system provided an online platform where the patients were able to submit their daily symptoms and vital signs, including temperature and blood oxygen saturation. With the convenient platform, the patient has recovered while being observed in their home. If their condition deteriorated, physicians would be notified and appropriate care would be provided. In addition, the system has shown that managing mild COVID-19 symptoms at home is a feasible option \cite{annis2020rapid}.

Users - patients and healthcare providers initially read the digits directly from the pulse oximeter display and upload the value to the healthcare server. Regarding this process, the users require a certain amount of time for correct typing and are prone to make an error input in their recording, causing a misinterpretation and a failure to detect a critical situation.

Although several studies attempted to extract the value of the vital sign from the image using an object character recognition (OCR) technique, which can read inputs and extract important information, such as text, from scanned documents or images \cite{hamad2016detailed}, this technology required more considerable expenses to access. For example, commercially available cloud-based OCR services, such as Amazon Rekognition and Google Vision API, cost \$1.0 - \$1.5 per one thousand image predictions, respectively. Supposing that COVID-19 patients are monitored twice daily on a nationwide level, it may cost as much as \$9000 a month. 

To overcome these challenges, we proposed a framework called \textit{PACMAN} (Pandemic Accelerated Human-Machine Collaboration) that can detect pulse oximeter display images and read in a real-world situation based on low-resource deep learning-based object detection.
In summary, The main contributions of this paper are as follows:
\begin{itemize}
    \item We promoted an online scalable crowdsourcing platform \footnote{\label{web:pacman}https://www.pacman-covid19.com/} that engaged the general public during the pandemic and allowed us to obtain pulse oximeter display dataset  in real-world situations as the input image into deep-learning based object detection models.

    \item The proposed PACMAN framework, an implemented digit auto-orientation and clustering algorithm on YOLOv5 that outperformed the state-of-the-art approach, achieved digit detection and digit reading using a crowdsourced dataset with varying quality and alignment. 

    \item PACMAN framework has currently deployed in real-world settings by integrating into a remote patient monitoring system over ten nationwide hospitals in Thailand.
\end{itemize} 
The rest of this article is structured as follows.  Section II reviews recent research on medical device digit recognition and object detection models. Section III explains the proposed framework. 
Section IV describes the experimental results. Section V discusses the results. Section VI shows the clinical deployment and future work. Finally, Section VII concludes this work.

\section{Related work}
This section outlines the development of digit recognition on medical device displays. Then, we conclude the limitations of the current research and provide appropriate approaches to rectify the constraint.

\subsection{Digits recognition on medical device displays}
Beginning with traditional methods, Morris et al. \cite{morris2006clearspeech} developed Clearspeech, a display reader for visually impaired people, by extending Eigenimage recognition to detect digital display numbers.
Nearly 90\% of evaluated displays are accurate in properly lighted.
Tsiktsiris et al. \cite{8956283} also proposed an OCR method to recognize the digit on a home blood pressure monitoring. The algorithm achieved a 96.2 \% accuracy rate. However, OCR  required additional implementations for eliminating artifacts and noise from the captured image to improve the performance of the approach.

In recent years, modern techniques for digit recognition on seven-segment displays have found application in a wide variety of machine learning-based applications.
In 2018, Shenoy et al. \cite{Shenoy2017} developed a smartphone application that could recognize digits from medical devices. Computer vision-based feature extraction such as Otsu's threshold, noise removal, and digit segmentation along with machine learning was used to transcript images of medical device displays into numbers.
The random forest model reached 98.2\% accuracy. However, users need manually crop the numbers, which is an additional effort, and images must be taken in the proper environment.

In 2019, Finnegan et al. \cite{finnegan2019automated} developed an automatic approach for detecting seven-segment digits in medical devices images with LCDs, precisely blood glucose meters and blood pressure monitors. The multi-layer perceptron (MLP) was used as a classifier and achieved 93\% accuracy on digits recognition. However, this approach used specific model of medical device and relies on some suitable environmental conditions.

In 2021, Kulkarni et al. \cite{kulkarni2021cnn} developed a transcription digits model of blood pressure from mobile phone cameras. They used the OpenCV library to preprocess the substantial variation in the appearance of the BP monitor images caused by orientation, zooming, and environmental conditions. Afterward, a convolutional neural network (CNN) was trained to transform LCD images into numeric values. This approach achieved 91\% accuracy, higher than Google Vision API and Tesseract OCR method but still has poor performance on low-resolution and noisy images and requires a large number of images to be used as a training data. In addition, this approach also requires a specific home blood pressure monitoring model with a user guidance interface in order to obtain high accuracy in the digit recognition task.

The aforementioned studies were conducted using only a frontal view of the display. 
Although some studies attempted to perform under varied viewpoints and environmental settings, they all used the same medical device model especially home blood pressure monitoring even there are numerous medical devices in daily life that have distinct digits alignment patterns, colors, and display styles. Consequently, the current recommended solution may not be applicable in remote health monitoring during COVID-19 pandemic.

\subsection{Object detection models}
\label{sec:object_detection}
As on-screen digits detection could also be counted as an object detection task, we explored multiple approaches to find the most suitable approach to rectify our problem of finding reliable and resource-efficient on-screen digits detection on the production system within our computational-power resources and crowdsourced image constraints.

You Only Look Once (YOLO) \cite{yolov1} and its variants \cite{yolov2, yolov4, yolov5}, is another family of high-throughput object detection algorithms that had proven its performance in the production environment. 
Being high-throughput and proven method is enough to be one of the candidate in our study.
Herein, we explored its improvement, YOLOv4, which enhanced from its predecessor YOLOv3 by a novel CSPDarknet53 \cite{cspnet} backbone and the introduction of spatial pyramid pooling to increase the model receptive field. Its improvement of the model and the information flow in the network leads to more accurate localization of the object. 
Scaled YOLOv4 \cite{scaled-yolov4} is an improvement to YOLOv4 using the model scaling method. By achieving higher runtime throughput, we chose to explore Scaled YOLOv4 further in our study. YOLOv5 is an enhanced version of YOLOv4 that incorporates mosaic augmentation into the data augmentation procedure to aid in the detection of small objects and an auto-learning bounding box mechanism in order to leverage manual anchor box size tuning. Additionally, YOLOv5 is substantially lighter than YOLOv4, which may be a more appropriate solution to our resource constraint issue.

You Only Learn One Representation (YOLOR) \cite{yolor}, a modified version of YOLO, is also a considerable object detector solutions. The primary distinction between YOLOR and YOLO is that YOLOR includes both implicit and explicit knowledge, which can enhance the model's recognition speed and accuracy. This lightweight architecture model may be able to solve our proposed problem in a limited resource setting while still providing reliable accuracy.

Another series of algorithms, Visual Transformer (ViT) \cite{vit} and its variants, may give a state-of-the-art prediction performance on our task. However, it is well known that this group of algorithms requires enormous computational resources, exceeding our limit. In accordance with the defined resource constraint, we decided to not include the transformer-based methods in our approach. 

\section{The proposed framework}
\begin{figure}[]
  \centering
    \captionsetup{justification=centering}
    \includegraphics[width=\columnwidth]{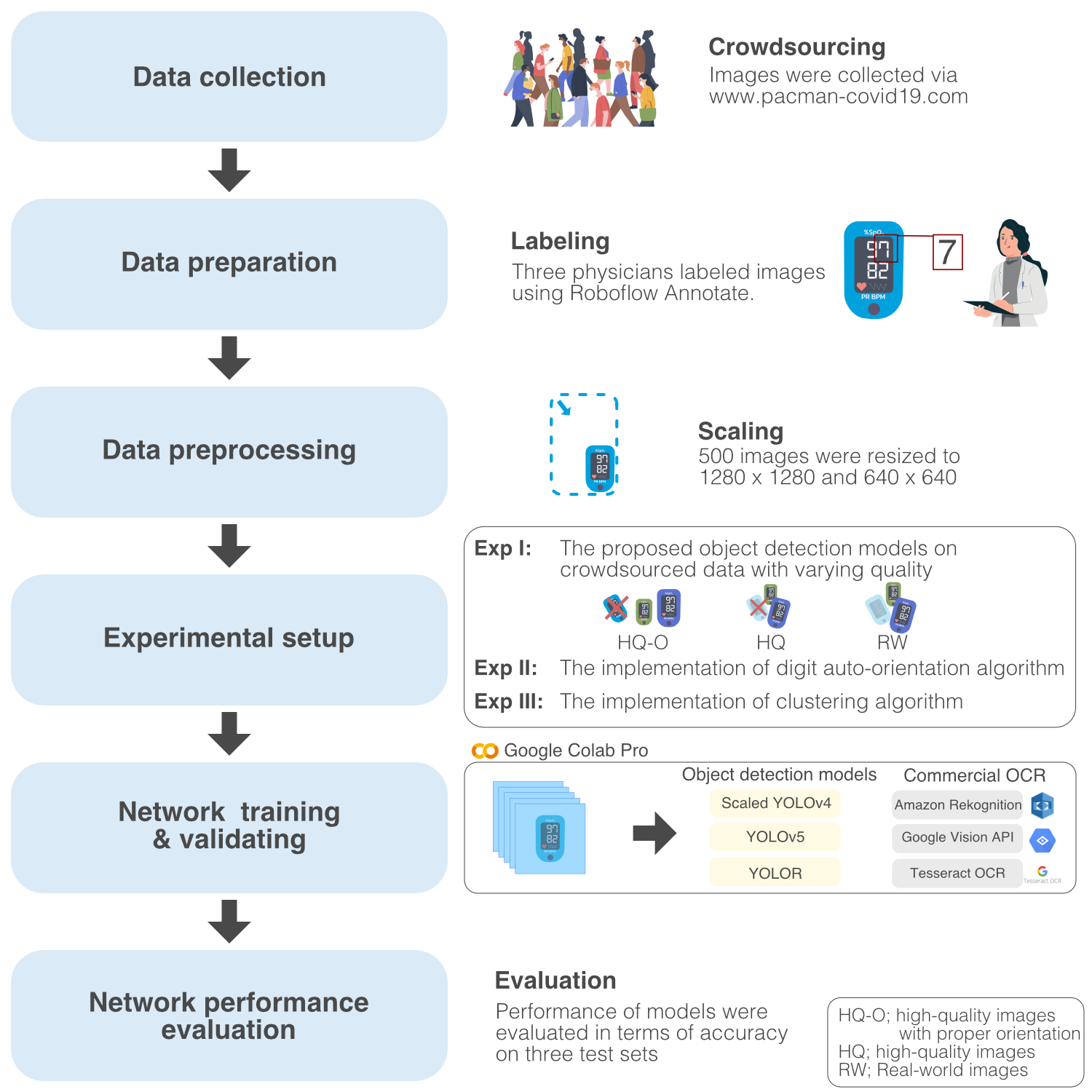}
    \caption{PACMAN framework}
    \label{fig:framework}
\end{figure}

Based on a survey of relevant literature, many object detection models could be candidates for the production system, despite restricted time and resources. However, a comparison of the performance of these models in detecting and recognizing on-screen digits was not available. In addition, there were no images of pulse oximeters obtained by real-world users under varying environmental conditions. To tackle this challenge, we proposed a novel PACMAN framework in \autoref{fig:framework} that gives better results in a real-world situation with the low-resource setting. The six steps that comprise our framework are as follows: (i) crowdsourcing data collection, (ii) data preparation and annotation, (iii) data preprocessing, (iv) experimental setup, (v) network training and validating, and (vi) network performance evaluation. The following subsections give more details about the proposed framework.  

\subsection{Crowdsourcing data collection}
To obtain a real-world pulse oximeter display dataset, we initiated a nationwide crowdsourcing campaign called \enquote{Pandemic Accelerated Human-Machine Collaboration (PACMAN)} to collect pulse oximeter displays via our online data collection platform \textsuperscript{\ref{web:pacman}}. Additionally, we persuaded user acquisition by promoting our campaign and raising awareness via social media.
Participants from across the nation who were interested in our campaign were required to upload images of their pulse oximeter displays taken by smartphone camera to our website and manually enter the oxygen saturation and pulse rate values. To encourage users to complete the task, the system returned the interpretation of their oxygen saturation suggested by healthcare providers. Uploaded images and texts were encrypted and stored securely in the Amazon S3 cloud storage. We did not collect sensitive information, email addresses, or explicit personal identifiers from our users.
\begin{figure}[]
    \centering
    \subfloat[]{\includegraphics[width=0.2\columnwidth]{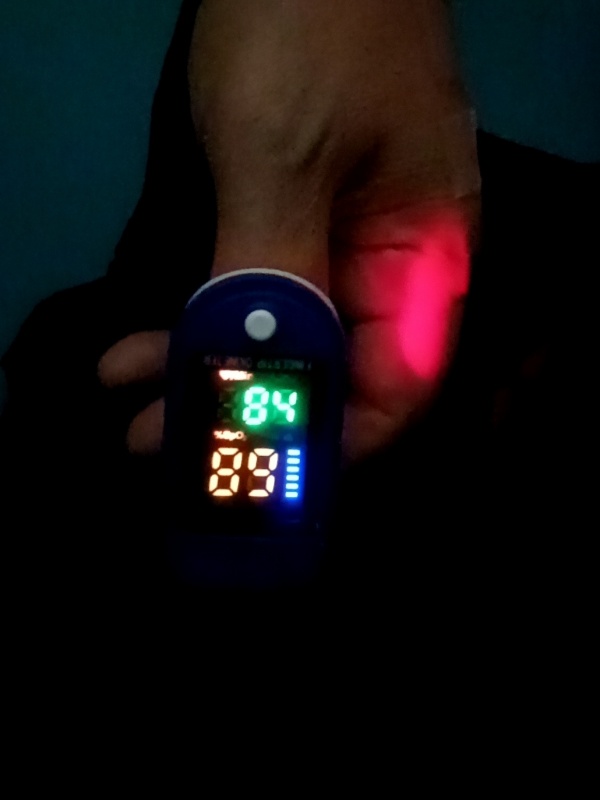}%
    \label{fig:low-light}}
    \hskip\baselineskip
    \subfloat[]{\includegraphics[width=0.2\columnwidth]{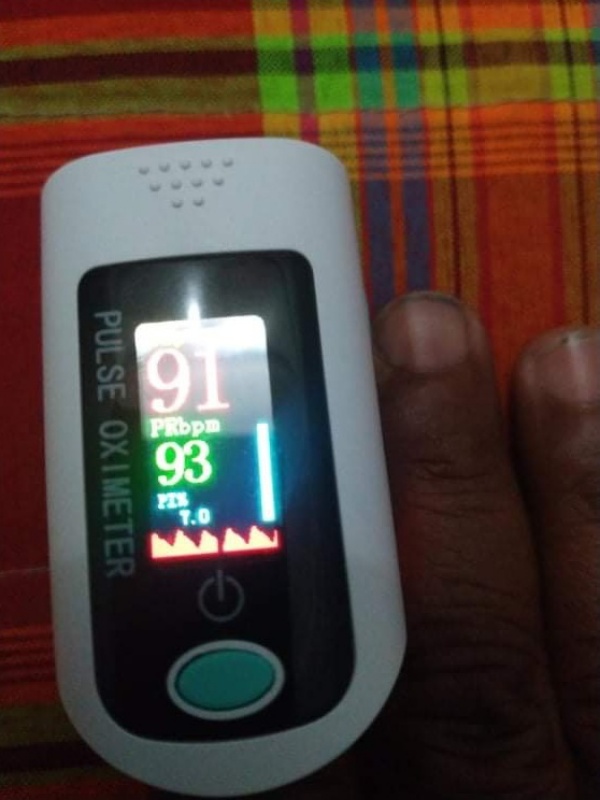}%
    \label{fig:overexposure}}
    \hskip\baselineskip
    \subfloat[]{\includegraphics[width=0.2\columnwidth]{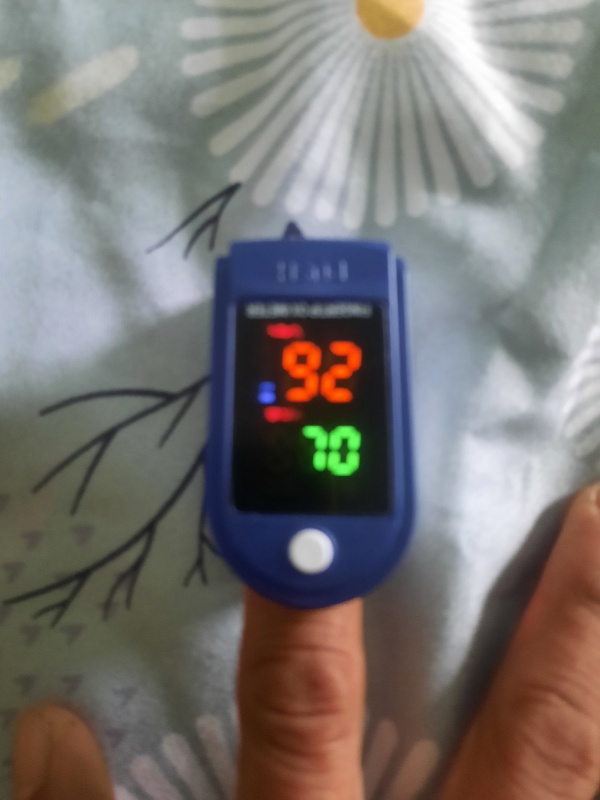}%
    \label{fig:lose-focus}}
    \hskip\baselineskip
    \subfloat[]{\includegraphics[width=0.2\columnwidth]{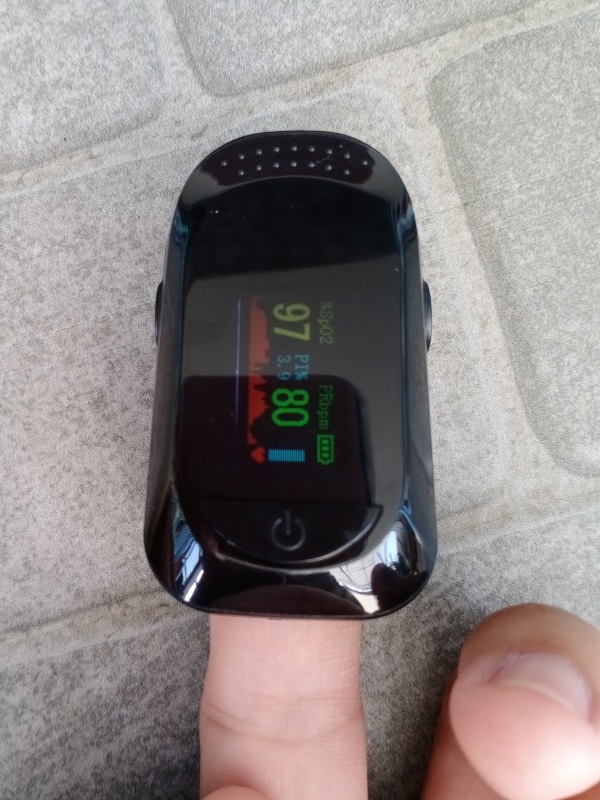} %
    \label{fig:screen-reflection}}
    \caption{\small Example images affected by the environmental conditions including the image of seven segment display with (a) loss of focus, (b) low-light setting, and the image of dot matrix display with (c) overexposure and (d) screen reflection.}
    \label{fig:env_display}
\end{figure}


Two months following the initiation of a crowdsourcing and health promotion campaign with the goal to educate individuals on how to interpret home pulse oximeters together with the uploading of pulse oximeter display images. All images were acquired in an uncontrolled environment: no constraints on background, lighting, perspective, rotations, occlusions, smartphone type, or pulse oximeter model. \autoref{fig:env_display}(a)-\autoref{fig:env_display}(d) illustrate four examples of images that had been impacted by environmental factors.

\subsection{Data preparation and annotation}
The next step was manually drawing boundary boxes on images using the annotation software Roboflow Annotate. 
Three physicians were assigned to annotate the dataset in order to reduce subjectivity. The first annotators drew boundary boxes and annotated all images. The second annotator reviewed and revised all annotated images to ensure their completeness. The third annotator along with the previous two reexamined all ambiguous images and reached a consensus on their interpretation.

\subsection{Data preprocessing}
\label{sec:preprocessing}
Given the considerable diversity in the resolution of pulse oximeter images based on users' smartphone cameras, preprocessing of the images was required before training models. As described in Section \ref{sec:exp1}, Scaled YOLOv4 \cite{scaled-yolov4}, YOLOv5 \cite{yolov5}, and YOLOR \cite{yolor} were used in this work. Considering image resolution, Scaled YOLOv4 can be trained on various image resolutions, including 512, 640, 896, and 1280 pixels. In contrast, YOLOv5 and YOLOR can only be trained on 640 and 1280 pixels. Therefore, all images were resized to 640 x 640 and 1280 x 1280 pixels, which was the common size required for each object detection model.

\subsection{Experimental setup}
\subsubsection{Experiment I: The proposed object detection models on crowdsourced data with varying quality}
\label{sec:exp1}

This experiment aimed to benchmark the performance of the state-of-the-art object detection models (Scaled YOLOv4 \cite{scaled-yolov4}, YOLOv5 \cite{yolov5}, and YOLOR \cite{yolor}), including three commercial OCR services (Amazon Rekognition, Google Vision API, and Tesseract OCR \cite{tesseract-ocr}) for digit detection and recognition in the different datasets to reduce the bias, making additive models in certain conditions. To achieve the experiment, we compared the proposed object detection models on the given image qualities as follows:

\begin{itemize}
    \item {\textit{High-quality image with proper orientation (HQ-O)}}
    The image dataset in HQ-O was under the lowest environmental impact (Research-graded image), which was no blurring, loss of focus, low-light setting, and screen reflection. Moreover, the display orientation of each image was manually rotated to either horizontal or vertical to minimize the effect of orientation.
   
    \item {\textit{ High-quality image (HQ)}}
    The image dataset was similar to the HQ-O dataset but was allowed incorrect display orientation to evaluate the impact of display orientation.
    
    \item {\textit{Real-world image (RW)}}
    The image dataset was obtained from a real-world situation without restriction.  As a result, low-quality images and incorrect display orientation images were included.
\end{itemize}

\subsubsection{Experiment II: The implementation of digit auto-orientation algorithm}
Experiment II was the extension of Experiment I by implementing the digit auto-orientation algorithm on the proposed models for experimental comparison across different image datasets, as described in Section \ref{sec:exp1}. The purpose of the experiment was to deal with a real situation where users occasionally provided images of pulse oximeters with incorrect display orientation. Thus, we implemented the digit auto-orientation algorithm to correct the image orientation, as described in Algorithm \autoref{algo:pipeline} (lines 3-8). Briefly, the algorithm rotated the original image with four different angles (0, 90, 180, and 270) before collecting the confidence values of all recognized digits from each inference. The angle with the highest median of confidence values among the recognized digits was chosen.
We assume that if the image orientation is proper, all digits must be detected with a higher median of confidence values than images with improper orientation.

\subsubsection{Experiment III: The implementation of clustering algorithm}
Experiment III extended Experiment II by applying a clustering method after the proposed models. The experiment's objective was to extract oxygen saturation and pulse rate data from the cluster of recognized numbers on pulse oximeter. In brief, the K-mean clustering algorithm was applied to divide a set of detected numbers into two groups: oxygen saturation and pulse rate, as shown in Algorithm \autoref{algo:pipeline} (lines 9-13). 
\begin{figure}[]
    \centering
    \subfloat[\small]{\includegraphics[width=0.75\columnwidth]{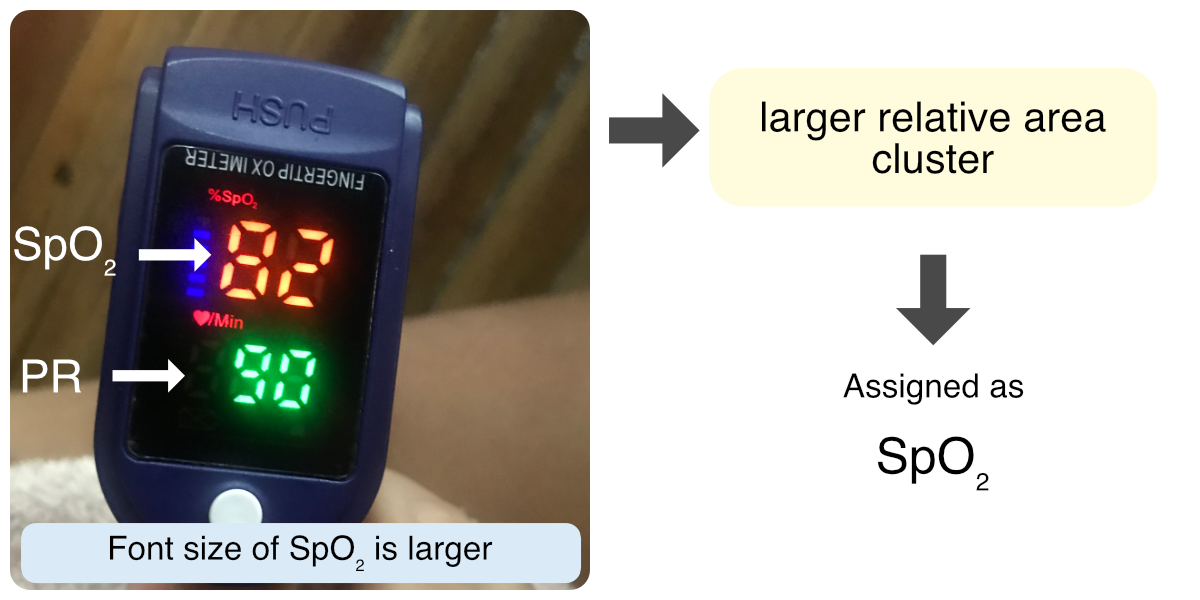}%
    \label{fig:same}}
    \\
    \subfloat[\small]{\includegraphics[width=0.75\columnwidth]{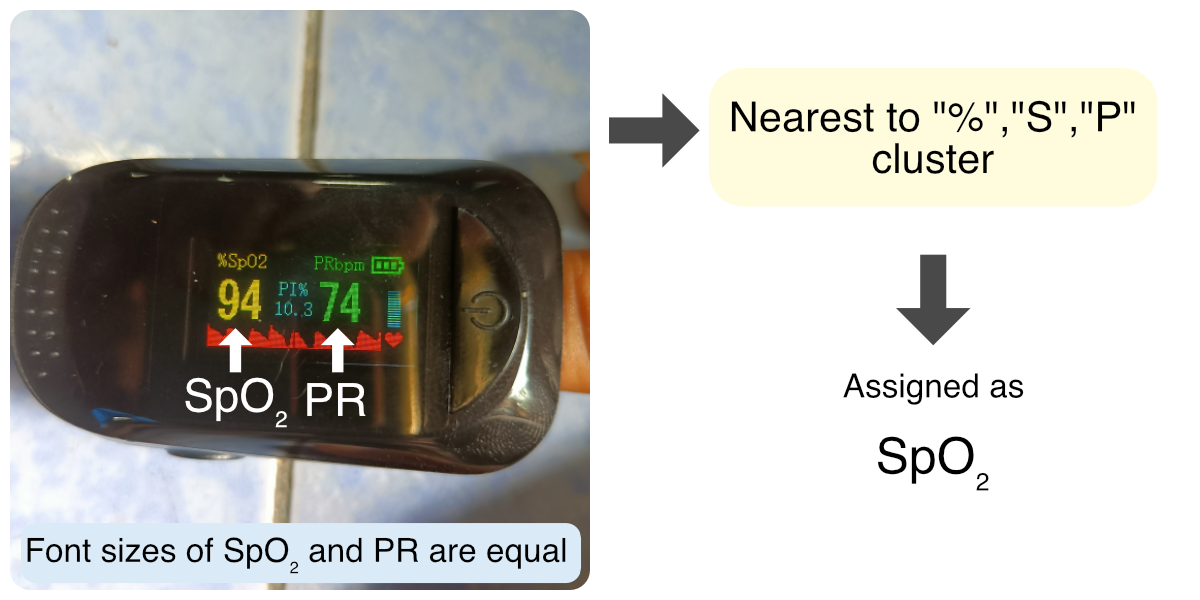}%
    \label{fig:diff}}
    \caption{\small Two main layouts of pulse oximeter displays: (a) font size of oxygen saturation is larger and (b) font sizes of oxygen saturation and pulse rate are equal}  
    \label{fig:display}
\end{figure}
Typically, there are two main layouts of pulse oximeter displays as shown in \autoref{fig:display}: 1) A font size of oxygen saturation is larger (\autoref{fig:display}(a)), and 2) font sizes of oxygen saturation and pulse rate are equal (\autoref{fig:display}(b)), as described in Algorithm \autoref{algo:pipeline} (lines 21-25). 
Accordingly, the larger relative area cluster was designated as oxygen saturation. For the second type, the distance between the first digits of each cluster and one of \enquote{\%}, \enquote{s} and \enquote{p} were calculated. Afterward, we assigned oxygen saturation to the closest cluster. Subsequently, the remaining cluster was assigned to the pulse rate, as described in Algorithm \autoref{algo:pipeline} (lines 21-25).

\subsection{Network training and validating}
\label{sec:training}
\subsubsection{Network training}
With limited time and resources, only Google Colab platform with Pro package (\$10/month) included a NVIDIA Tesla V100-SXM2 graphic card, 12.68GB of RAM, and 8 CPU cores were available at hand. To ensure a fair model comparison in Experiment I, transfer learning through pre-training and additional explicit regularization (weight decay, dropout, and Batch normalization) were not applied. However, we did not adjust the data augmentation techniques and regularization in each architecture's backbone. All models were trained using the training set presented in Section \ref{sec:validation} with the default hyperparameter configuration. A stochastic gradient descent (SGD) as a default optimizer of all three models was used to minimize the cross-entropy loss.
Each model was trained for a maximum of 500 epochs with early stopping to prevent that training period from exceeding 24 hours, as limited by Google Colab Pro package. Additionally, early stopping was also used to terminate training when the validation loss stopped improving. 

\algnewcommand\Or{\textbf{ or } }
\algnewcommand\Do{\textbf{ do } }
\algnewcommand\Continue{\textbf{ continue } }
\algrenewcommand\alglinenumber[1]{\tiny #1:}

\begin{algorithm}[t]
    \scriptsize
    \caption{Digit Auto-orientation and clustering}\label{ALGO1}
    \label{algo:pipeline}
    \textbf{Input:} $\textit{image}$, $\textit{digit\_inference}$, $\textit{sp\_inference}$  \\
    \textbf{Output:} [$SpO2$, $PR$]
    \begin{algorithmic}[1]
        \State $\textit{median\_of\_conf}$ = []
        \State $\textit{result\_list}$ = []
        \For {$\textit{rot}$ in [0, 90, 180, 270]}
            \State rotate image by $\textit{rot}$
            \State $\textit{digits\_list}$, $\textit{conf\_list}$ = digit\_inference($\textit{rotated\_image}$) 
            \State calculate median of all elements in $\textit{conf\_list}$ and insert into $\textit{median\_of\_conf}$
            \State insert $\textit{digits\_list}$ into $\textit{result\_list}$
        \EndFor
        \State sort $\textit{median\_of\_conf}$ and corresponding elements in $\textit{result\_list}$ in descending order 
        \For {$\textit{det}$ in $\textit{result\_list}$ }
            \If {$\textit{det}$ has 5 digits}
                \State $\textit{k}$ = 2
            \ElsIf{$\textit{det}$ has more than or equal to 6 digits}
                \State $\textit{k}$ = 3
            \EndIf
            \State $\textit{label}$ = Kmean($\textit{k}$, initial\_pos)
            \State concatenate digits follow the generated $\textit{label}$
            \State $\textit{list\_sp}$ = sp\_inference(rotated\_image) 
        \State (*)
            \If{$\textit{list\_sp}$ is empty} 
                \State $SpO2 \gets$ the cluster with bigger relative area
                \State $PR \gets$ another cluster
            \Else 
                \State $\textit{distance}$ = Euclidean distance between the leftmost digit of each cluster, and one of \%, s, p in $\textit{list\_sp}$
                \State sort $\textit{distance}$ in ascending order
                \State $SpO2 \gets$ nearest cluster
                \State $PR \gets$ farther cluster
            \EndIf
            \If {$SpO2$ $\not\in$ [70,100]\Or $PR$ $\not\in$ [40,300]} 
                \State Exclude the outlier element in each cluster by height of bounding box
                \State \Do (*)
            \EndIf 
        
            \If {$SpO2$ $\not\in$ [70,100]\Or $PR$ $\not\in$ [40,300]}
                \State \Continue     
            \EndIf
        \State \Return[$SpO2$, $PR$]
        \EndFor
    \end{algorithmic}
\end{algorithm}

\subsubsection{Network validating}
\label{sec:validation}
We obtained a total of 1,012 images from crowdsourced data \footnote{The crowdsourced data used to support the findings of this study is available upon request. All inquiries regarding this content should be directed to the corresponding author.}. We observed that there are two main types of displays for medical devices; (i) seven-segment (SSD) - a squared-off figure eight comprised of seven bar-shaped LED or LCD elements, and (ii) dot-matrix (DMD) - a high-quality display. For all of these, we cut off the oxygen saturation at 95 \% which normal oxygen saturation (N) is defined as $\geq$ 95\% \cite{louw2001accuracy}. Otherwise, the oxygen saturation level is considered low (L). Regarding the characteristics of the dataset, we classified them into four groups based on the display type and the oxygen saturation level, as shown in \autoref{tab:number_image}

\begin{table}[]
\centering
\caption{Characteristics of the crowdsourced dataset}
\label{tab:number_image}
\resizebox{1\columnwidth}{!}{%
    \begin{tabular}{lc}
\toprule[0.2em]
\multicolumn{1}{c}{Group}  & Crowdsourced dataset
\\ \midrule[0.1em]
SSD with normal oxygen saturation (SSD-N) & 238   \\
SSD with low oxygen saturation (SSD-L)    & 125   \\
DMD with normal oxygen saturation (DMD-N) & 437   \\
DMD with low oxygen saturation (DMD-L)    & 212  \\ 
\bottomrule[0.2em]
\\
\multicolumn{2}{l}{SSD; Seven segment display, DMD; Dot-matrix display}\\
\end{tabular}

    }
\end{table}

As we see, the number of images in each group is not equal, suggesting that our crowdsourced data are imbalanced. To address the problem with the unbalanced dataset, random undersampling \cite{kotsiantis2006handling} was used to ensure a balanced dataset in each group. Given that we had a minimum dataset of SSD-L (125 images) in the crowdsourced dataset, we randomly selected 125 images per group. Finally, we had 500 images for the training set which was further split into a new training set (80\%) and validating set (20\%) by using five-fold cross-validation.  

All trained models were finally validated with three external unseen datasets based on different image qualities and alignments, as illustrated in Section \ref{sec:exp1} to assess the performance of the proposed models in a variety of circumstances. The images were collected from real-world cases in community hospitals that use CHIVID \footnote{AI-Driven \textbf{C}ommunity/\textbf{H}ome \textbf{I}solation-Based Electronic Health Record during CO\textbf{VID}-19 pandemics or CHIVID is a remote patient monitoring platform used by many hospitals in Thailand} to monitor patients with mild COVID-19. Each external unseen dataset included 200 images, with 50 images in each group (SSD-N, SSD-L, DMD-N, and DMD-L). 

Note that, all training, validating, and testing sets were resized in different resolutions, as described in Section \ref{sec:preprocessing} before feeding into the proposed models to address the varieties of resolution.

\subsection{Network performance evaluation}
The model's performance was assessed by a mean average precision (mAP) metric with a specific Intersection over Union (IoU) threshold to compare the results. IoU can be computed as:

 \begin{equation}
 \label{eq:iou}
 IoU_{i} = \frac{P_{i} \cap G_{i}}{P_{i} \cup G_{i}}
 \end{equation}
 
Where $P_{i}$ indicates the bounding box i predicted by each model as described in Section \ref{sec:object_detection} and $G_{i}$ denotes the comparable bounding box in the ground truth. If the IoU of a predicted bounding box exceeds a 0.5 IoU threshold, the sample is classified as a true positive (TP). Otherwise, it is classified as a false positive (FP). In addition, regions that are in ground truth but not detected by models are classified as false negative (FN) samples. 

With these three components, we could calculate precision (P) and Recall (R). The average precision (AP) is defined as the area under the precision-recall curve, which illustrates the trade-off between precision and recall at various thresholds.
The mean average precision (mAP), which is the average of each class-specific AP score, is used to compare the performance of the models on validation sets. 

 \begin{equation}
 \label{eq:ap}
  AP = \int_{0}^{1} P(R) \,dR 
 \end{equation}
 
Note that, commercial object detection services, such as Amazon Rekognition and Google Vision API, did not provide the bounding boxes for each digit on their outputs. Thus, IoU and mAP were not sufficient to be used for the comparison.
Therefore, in Experiment I-III, we used the accuracy score to determine which models outperform.
\newline

In Experiment I and II, We defined $Pred_{i}$ as a set of predicted numbers from image ${i}$ and $Ground_{i}$ defined as a set of ground truth numbers in image ${i}$. The sample ${i}$ was recognized correctly when set $Pred_{i}$ equaled to set $Ground_{i}$ as described in \autoref{eq:correct12}. 

\begin{equation}
\label{eq:correct12}
  C_{i} =
    \begin{cases}
      1 & \text{if $Pred_{i} = Ground_{i}$}\\
      0 & \text{if $Pred_{i} \neq Ground_{i}$ }
    \end{cases}       
\end{equation}

However, the prediction was defined differently in Experiment III which aimed to extract the value of oxygen saturation and pulse rate. Therefore, $C_{i}$ was equal to 1 when both predicted SpO\textsubscript{2} and predicted PR were equal to their ground truth values. Other events were defined as 0. Consequently, the overall accuracy was defined as:
\begin{equation}
 \label{eq:accuracy}
 Accuracy = \frac{\sum C_{i}}{N} \times 100 \%
\end{equation}

Where $N$ was the number of test set images.

\section{Experimental results}
\begin{figure*}[]
    \centering
    \subfloat[\small]{\includegraphics[width=0.25\textwidth]{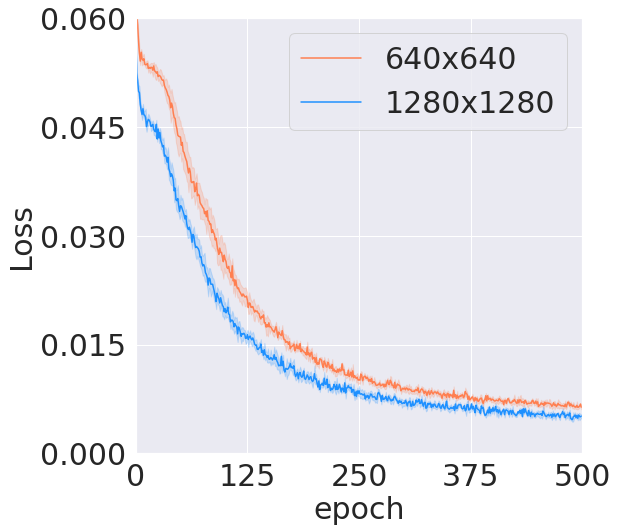}%
    \label{fig:loss-yolov4s-640}}
    \hskip\baselineskip
    \subfloat[\small]{\includegraphics[width=0.25\textwidth]{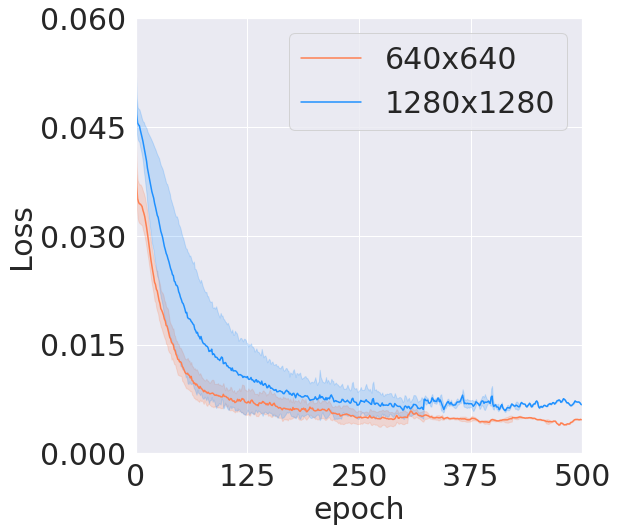}%
    \label{fig:loss-yolov5-640}}
    \hskip\baselineskip
    \subfloat[\small]{\includegraphics[width=0.25\textwidth]{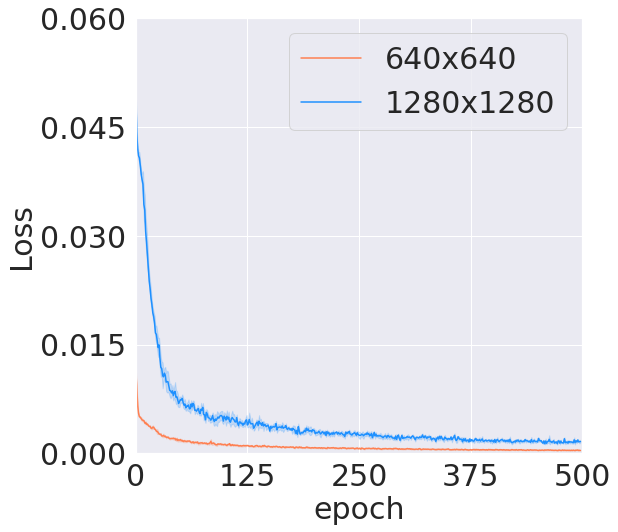}%
    \label{fig:loss-yolor-640}}
    \caption{\small Validation averaged losses with standard deviation of the models on 640 x 640 (orange line) and 1280 x 1280 (blue line) resolutions: (a) YOLOv4, (b) YOLOv5 and (c) YOLOR.}  
    \label{fig:train-val-loss}
\end{figure*}

\subsection{Experiment I: The proposed object detection models on crowdsourced data with varying quality and alignment}

\begin{table}[b]
\centering
\small
\caption{Digit detection models on pulse oximeter display dataset comparisons in terms of mAP percentage on training-validation set}
\label{tab:train}
\resizebox{1\columnwidth}{!}{%
    \begin{tabular}{@{}ccccc@{}}
\toprule[0.2em]
\multicolumn{1}{l}{\textbf{Resolution}} &\textbf{Model} &\textbf{\# Params (M)} & \textbf{mAP\text{@}50 (mean $\pm$ SD)} & \textbf{Inference (ms)}
\\ \midrule[0.1em]
    
    \multirow{3}{*}{640 x 640} 
    
    &Scaled YOLOv4 
    &52.5  
    &0.958 $\pm$ 0.013 
    &28.0 $\pm$ 3.4	\\
     
    &YOLOv5 
    &83.3 
    &\textbf{0.967 $\pm$ 0.012}	
    &20.8 $\pm$ 0.9\\
    
    &YOLOR 
    &37.2	
    &0.943 $\pm$ 0.022
    &\textbf{17.2 $\pm$ 0.6} \\

    \midrule[0.1em]
     
    \multirow{3}{*}{1280 x 1280} 
    
    &Scaled YOLOv4 
    &70.3 
    &0.953 $\pm$ 0.018 
    &\textbf{36.0 $\pm$ 3.7}\\
     
    &YOLOv5 
    &140.1
    &\textbf{0.969 $\pm$ 0.017}	
    &59.1 $\pm$ 4.8\\
    
    &YOLOR 
    &151.6
    &0.954 $\pm$ 0.015	
    &54.3 $\pm$ 4.4 \\

    \bottomrule[0.2em]\\

\end{tabular}

    }
\end{table}
As indicated in \autoref{fig:train-val-loss}, it was observed that YOLOR converged more rapidly than other models. Despite the fact that, YOLOv5 had a large confidence interval for training and validation loss at both resolutions, it achieved highest performance in terms of mean average precision as shown in \autoref{tab:train}.  
\autoref{tab:experiment123} provided an overview of the comparative results that examined the model's performance in different settings. All models performed better on the HQ-O test set, which consisted of images with low environmental impact. 
In comparison to other models, the performance of YOLOv5 was superior on both resolutions. However, inference time of YOLOv5 was slower than scaled YOLOv4. 
YOLOv5 on 640x640 resolutions achieved the highest level of accuracy on all test sets, particularly the HQ-O test set. In addition, Its performance was slightly improved by approximately 4\%, 2\% and 1\% for HQ-O, HQ, and RW test sets, respectively, at 1280 x 1280 resolutions but it was slower speed inference time than 640 x 640 resolutions.
Compared to commercial OCR service, The performance of YOLOv5 was remarkably higher than that of Amazon Rekognition and Google Vision API while Tessaract OCR produced an unsatisfactory result due to a transcription error. 

Based on the results, YOLOv5 was the best model performance for digit detection and recognition tasks of all quality and alignment datasets, notably for the proper image dataset, at both resolutions. %

\begin{table*}
\centering
\large
\caption{The model performances based on the experimental designs are reported by the percentage of accuracy (mean $\pm$ SD) for digit recognition in the different datasets.}

\label{tab:experiment123}
\resizebox{0.75\textwidth}{!}{%
    \begin{tabular}{@{}cccccccc@{}}
\toprule[0.2em]

\textbf{Experiment} & \textbf{Resolution} & \textbf{Model} & \textbf{Algorithm} & \textbf{HQ-O(\%)} & \textbf{HQ(\%)} & \textbf{RW(\%)} & \textbf{Inference(ms)}
\\ \midrule[0.1em]

    \multirow{9}{*}{Experiment I} 
    &\multirow{3}{*}{640 x 640} 
    &Scaled YOLOv4 
    &none
    &84.0 $\pm$ 6.3	
    &54.5 $\pm$ 11.8 
    &51.5 $\pm$ 15.9	
    &\textbf{18.5 $\pm$ 3.3}
    \\
    &
    &YOLOv5 
    &none
    &\textbf{90.5 $\pm$ 1.9}	
    &\textbf{60.0 $\pm$ 10.2}	
    &\textbf{58.5 $\pm$ 14.4}
    &19.2 $\pm$ 0.8	
    \\
    &
    &YOLOR 
    &none
    &89.0 $\pm$ 3.5	
    &58.5 $\pm$ 9.1	
    &54.5 $\pm$ 14.7  
    &28.3 $\pm$ 2.8  
    \\

    \cline{2-8}
     
    &\multirow{3}{*}{1280 x 1280} 
    &Scaled YOLOv4 
    &none
    &86.5 $\pm$ 1.0	
    &55.5 $\pm$ 8.7 
    &52.5 $\pm$ 12.5	
    &57.4 $\pm$ 2.5  
    \\
    &
    &YOLOv5 
    &none
    &\textbf{94.5 $\pm$ 4.7 }
    &\textbf{62.0 $\pm$ 9.9}	
    &\textbf{59.5 $\pm$ 15.4}
    &63.3 $\pm$ 5.5 	
    \\
    &
    &YOLOR 
    &none
    &87.0 $\pm$ 3.5	
    &55.0 $\pm$ 10.9	
    &51.5 $\pm$ 15.4  
    &\textbf{48.8 $\pm$ 1.3}  
    \\

    \cline{2-8}

    &\multirow{3}{*}{Raw} 
    &Amazon Rekognition 
    &none
    &49.5 $\pm$ 32.9	
    &45.0 $\pm$ 33.1 	
    &39.0 $\pm$ 34.2	
    &- 
    \\
    &
    &Google Vision API   
    &none
    &25.5 $\pm$ 23.0	
    &27.5 $\pm$ 21.5 	
    &23.5 $\pm$ 23.9	
    &-  
    \\
    &
    &Tesseract OCR 
    &none
    &0	
    &0 
    &0	
    &-\\
     
    \midrule[0.1em]

    \multirow{6}{*}{Experiment II}
    &\multirow{3}{*}{640 x 640}     
    &Scaled YOLOv4 
    & AO
    &84.0 $\pm$ 6.3	
    &84.0 $\pm$ 6.3 
    &73.5 $\pm$ 10.0	
    &-
    \\
    &
    &YOLOv5
    & AO
    &\textbf{91.0 $\pm$ 2.6}  	
    &\textbf{91.0 $\pm$ 2.6} 
    &\textbf{82.5 $\pm$ 9.1} 
    &-
    \\
    &
    &YOLOR 
    & AO
    &89.5 $\pm$ 3.0
    &89.5 $\pm$ 3.0
    &77.5 $\pm$ 12.0  
    &-
    \\
    \cline{2-8}
    
    &\multirow{3}{*}{1280 x 1280} 
    &Scaled YOLOv4 
    & AO
    &86.5 $\pm$ 1.0	
    &86.5 $\pm$ 1.0 
    &75.5 $\pm$ 10.4	
    &-
    \\
    &
    &YOLOv5 
    & AO
    &\textbf{94.5 $\pm$ 4.7} 	
    &\textbf{94.5 $\pm$ 4.7}
    &\textbf{86.5 $\pm$ 11.2} 
    &-
    \\
    &
    &YOLOR 
    & AO
    &87.0 $\pm$ 3.5 	
    &87.0 $\pm$ 3.5	
    &75.0 $\pm$ 14.5  
    &-
    \\
    \midrule[0.1em]

    \multirow{6}{*}{Experiment III}
    &\multirow{3}{*}{640 x 640} 
    &Scaled YOLOv4 
    & AO+Clustering
    &81.5 $\pm$ 6.0	
    &81.5 $\pm$ 6.0 
    &71.0 $\pm$ 14.1	
    &-
    \\
    &
    &YOLOv5 
    & AO+Clustering
    &85.0 $\pm$ 4.2  	
    &85.0 $\pm$ 4.2 
    &\textbf{77.0 $\pm$ 12.5} 
    &-
    \\
    &
    &YOLOR 
    & AO+Clustering
    &\textbf{87.5 $\pm$ 4.4}
    &\textbf{87.5 $\pm$ 4.4}
    &75.0 $\pm$ 15.4  
    &-
    \\

    \cline{2-8}
    
    &\multirow{3}{*}{1280 x 1280} 
    &Scaled YOLOv4 
    & AO+Clustering
    &84.5 $\pm$ 4.7	
    &84.5 $\pm$ 4.7 
    &74.5 $\pm$ 12.8	
    &-
    \\
    & 
    &YOLOv5 
    & AO+Clustering
    &\textbf{89.5 $\pm$ 8.1} 	
    &\textbf{89.5 $\pm$ 8.1}
    &\textbf{81.0 $\pm$ 16.4} 
    &-
    \\
    &
    &YOLOR 
    & AO+Clustering
    &86.0 $\pm$ 4.3 	
    &86.0 $\pm$ 4.3	
    &74.5 $\pm$ 15.4  
    &-
    \\
    \bottomrule[0.2em]
    \\
    \multicolumn{8}{l}{AO; a digit auto-orientation algorithm, Clustering; a clustering algorithm }
\end{tabular}

    }
\end{table*}

\begin{table*}
\centering
\footnotesize
\caption{Comparison with existing studies on medical device display dataset}
\label{tab:compare_dataset}
\resizebox{0.9\textwidth}{!}{%
\newcommand{\etal}{\textit{et al.}}

\begin{tabular}{ccccccc}
\hline

\toprule[0.1em]
\multicolumn{1}{c}{\textbf{Work}} &\textbf{Model} &\textbf{Medical Device Display} & \textbf{Images} & \textbf{Accuracy} & \textbf{Strength} & \textbf{Limitations}
\\ \midrule[0.1em]

Shenoy \etal \cite{Shenoy2017} 
& Random Forest 
& \thead{BP monitor, weight scale,\\ and glucose meter}
& N/A
& 98.2\%  
& \thead{good performance \\on sharp images}
& \thead{poor performance \\on blurry images} \\ \hline

Tsiktsiris \etal \cite{8956283}      
& OCR 
& BP monitor                           
& N/A
& 96.2\% 
& No requiring training set
& Noise from captured image\\ \hline

Finnegan \etal \cite{finnegan2019automated}
& MLP classifier  
& \thead{BP monitor, \\and glucose meter}    
& 300
& 93.0\%   
& implementation of digit location algorithm
& \thead{poor performance \\for clustering the number} \\ \hline

\multirow{3}{*}{Kulkarni \etal \cite{kulkarni2021cnn}} 

& Tesseract   
& BP monitor                           
& 8,192
& 6.7 - 20.2\%    
& open-source
& errors in digit recognition process   \\

& Google Vision API   
& BP monitor                           
& 8,192
& 23.0 - 43.2\%    
& commercially available
& errors in digit recognition process  \\

& CNN
& BP monitor                           
& 8,192
& 61.7 - 91.1\%    
& \thead{good performance on\\high quality images}
& \thead{poor performance on\\low quality images}  \\ \hline

PACMAN framework
& \thead{YOLOv5 with \\AO+Clustering}
& Pulse Oximeter                           
& 500
& \textbf{81.0 - 89.5\%}    
& \thead{good performance\\on real-world dataset\\ with extracted SpO\textsubscript{2} and PR values}
& \thead{require more testing on\\other medical devices}\\ 

\bottomrule[0.1em]\\
\multicolumn{7}{l}{SpO\textsubscript{2}; Oxygen saturation, PR; Pulse rate}\\
\multicolumn{7}{l}{AO; a digit auto-orientation algorithm, Clustering; a clustering algorithm }
\end{tabular}

    }
\end{table*}

\subsection{Experiment II: The implementation of digit auto-orientation algorithm}
In accordance with Experiment I, the accuracy of all models' (scaled YOLOv4, YOLOv5, and YOLOR) on the HQ and RW test sets was inferior to the accuracy on the HQ-O test set. The result implied that improper image orientation could diminish model performance. Therefore, the digit auto-orientation algorithm was applied to each model with the best weight from Experiment I. After implementing the digit auto-orientation algorithm, at both resolutions, the accuracy of all models on the HQ and RW dataset were notably enhanced compared to without the implementation, as shown in \autoref{tab:experiment123}. The result suggested that the implemented digit auto-orientation algorithm ameliorated the accuracy of all models on improper image orientation.

\subsection{Experiment III: The implementation of clustering algorithm}
In a real-world setting, it is necessary to read entire oxygen saturation and pulse rate. Although models from Experiments I and II could accurately detect and recognize digits, they are unable to obtain oxygen saturation and pulse rate values directly. Therefore, we integrated the clustering algorithm into Experiment II to read the oxygen saturation and pulse rate values. The performances of each model sacrificed a small amount of accuracy but remained superior to that of the model without any additional algorithm (Experiment I), as illustrated in \autoref{tab:experiment123}.

\section{Discussion}
In this section, we discussed the results of three experiments that contributed to the completion of our framework, resulting in the ability to detect digit recognition and read the value from pulse oximeters. We also compared this work with the existing studies. Finally, we extend the deployment of our framework in the clinical context. 

As demonstrated in Experiment I, we conducted the object detection model on crowdsourced data with different image qualities. From the result in \autoref{tab:experiment123}, YOLOv5 was the best performing model against the other models on the different datasets, notably correct image orientation in both resolutions. However, the speed of inference time was still lower than others. It relied on a larger parameter during the training phase, leading to a better performance with low computational speed. Furthermore, the performance of YOLOv5 with 1280 x 1280 input size provided higher accuracy but spent longer inference time than 640 x 640 due to a larger scale with many parameters optimizing. Accordingly, we found the trade-off between accuracy and inference time. Although the fast speed inference time is preferable in real-time use, the accuracy performance for digit recognition is superiorly important to time processing. Thus, YOLOv5 with 1280 x 1280 input size was considered the appropriate model for digit detection across image qualities and alignments that high accuracy was dominant in the proper image orientation dataset. 

To enhance the accuracy performance of the model on improper image orientation under uncontrolled environments with various alignments, we implemented the auto digit orientation algorithm (Experiment II). The implementation automatically detected image rotation with the estimate of the confidence score in different angles by ranking the highest score that indicated the increased probability of the image beings detected correctly. As a result, the accuracy of all models was considerably improved on the improper image orientation dataset whose performances were very close to the proper image orientation dataset. 

However, Experiment I and Experiment II were not applicable in clinical monitoring without digit interpretation. Therefore, we needed to further implement clustering algorithms in experiment III to separate the value between  blood oxygen level and pulse rate. According to the results, the accuracy performances of the models were slightly dropped off across all image datasets. It may result from device diversity in terms of formats and perspective view. However, the reduced accuracy was still acceptable.

So far as we know, this is the first study that introduced the complete framework for digit recognition and digit interpretation of pulse oximeters in real-world datasets. While the existing study mainly focused on home blood pressure monitor and their performance on digit recognition tasks, as shown in \autoref{tab:compare_dataset}. 

\begin{figure}[]
    \centering
    \subfloat[]{\includegraphics[width=0.25\columnwidth]{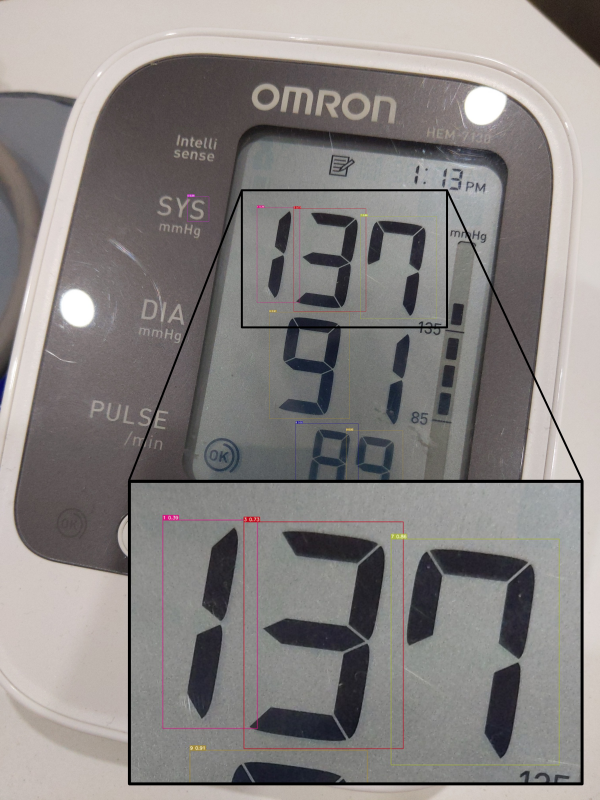}%
    \label{fig:map-yolov4s}}
    \hskip\baselineskip
    \subfloat[]{\includegraphics[width=0.25\columnwidth]{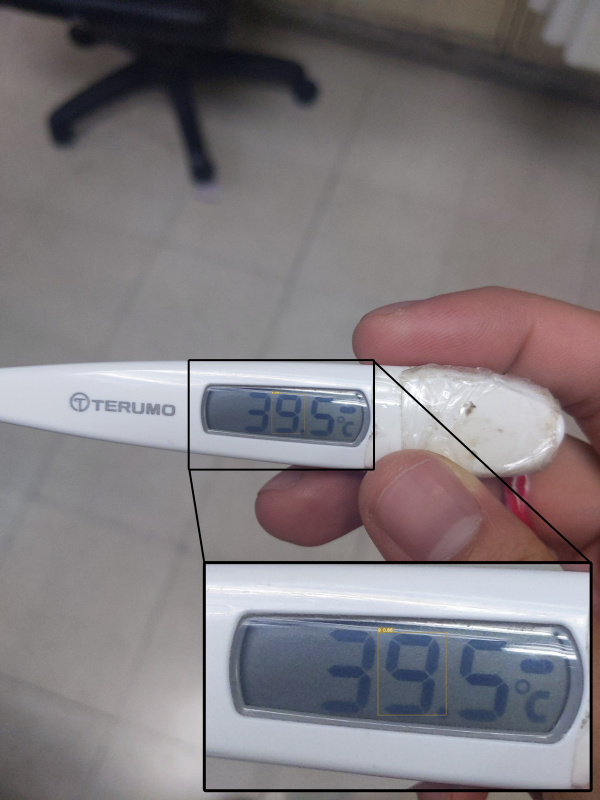}%
    \label{fig:map-yolov5}}
    \hskip\baselineskip
    \subfloat[]{\includegraphics[width=0.25\columnwidth]{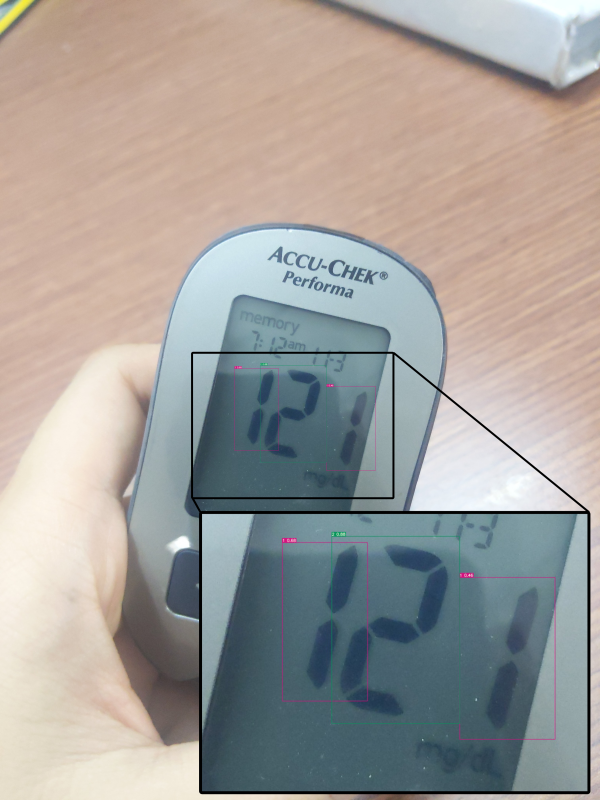}%
    \label{fig:map-yolor}}
    \caption[generalized to other medical devices ]
        {\small 
        YOLOv5 on pulse oximeter dataset was generalizable to various medical devices including (a) a blood pressure monitor, (b) a digital thermometer and (c) a glucose meter.}  
    \label{fig:generalized}
\end{figure}

We recently deployed the framework as two server instances. One is dedicated to the front-end site serving \textsuperscript{\ref{web:pacman}} and another for machine learning model service serving. For the deployment, the trained model was compiled into ONNX format, followed by graph optimization and model pruning to increase inference speed on the production server. Ten hospitals nationwide adopted this system for trials during the COVID-19 pandemic, responding favorably to the entire solution, notably usability and validation. Moreover, by utilizing our proposed framework, healthcare providers can effortlessly gather and enter other medical data, particularly vital signs, into Electronic Health Record (EHR), as depicted in \autoref{fig:generalized}. Consequently, our deployment strongly showed the highly usable for clinical settings. 

\section{Conclusion}
This study proposed a framework called PACMAN to detect and read digits of oxygen saturation and pulse rate on real-world on-screen images of the pulse oximeter. 
With limited time and resources, we compared three current states-of-the-art object detectors by evaluating them in real-world images crowdsourced from actual end users. 
The digit auto-orientation algorithm was implemented on the models to improve the detection and recognition performance. 
We then applied the clustering algorithm to extract oxygen saturation and pulse rate value from recognized digits. 
YOLOv5 with digit auto-orientation and clustering algorithm, as the best performing model, reached 81.0 - 89.5 \% of accuracy, which outperforms scaled YOLOv4, YOLOR and commercially available OCR services (i.e., Amazon Rekognition and Google Vision API).
Consequently, we integrated the PACMAN framework into a remote health monitoring system used by many hospitals across the country to validate its effectiveness and receive feedback from healthcare professionals. In future work, a lager number of image from real-world scenarios in healthcare service could be gathered to obtain a more robust accuracy of the model's performance.

\bibliographystyle{IEEEtran}
\bibliography{references}

\end{document}